\documentclass[prl,aps,onecolumn,showpacs]{revtex4}

\usepackage{psfrag}
\usepackage{graphicx}
\usepackage{graphics}
\usepackage{bm}
\usepackage{color}
\usepackage{centernot}
\usepackage{verbatim,ulem}
\usepackage{epstopdf}
\newcommand{\be}{\begin{equation}}
\newcommand{\ee}{\end{equation}}
\newcommand{\ba}{\begin{eqnarray}}
\newcommand{\ea}{\end{eqnarray}}

\newcommand{\dcom}[1]{}
\newcommand{\dnote}[1]{}

\newcommand{\nn}{\nonumber}

\newcommand{\gsim}{\raise.3ex\hbox{$>$\kern-.75em\lower1ex\hbox{$\sim$}}}
\newcommand{\lsim}{\raise.3ex\hbox{$<$\kern-.75em\lower1ex\hbox{$\sim$}}}

\begin{document}

\renewcommand{\thefootnote}{\fnsymbol{footnote}}


\renewcommand{\thefootnote}{\arabic{footnote}}
\setcounter{footnote}{0} \typeout{--- Main Text Start ---}

\title
{When are two fermions a simple boson?

 New Gross-Pitaevskii actions for cold Fermi condensates}
\author{Ray\ J.\ Rivers$^1$}
\author{D.A.~Steer$^2$}
\author{ Chi-Yong  Lin$^3$}
\author{Da-Shin Lee$^{3}$}
\author{D.J.~Weir$^4$}
\affiliation{$^1$ Blackett Laboratory, Imperial College\\
London SW7 2BZ, U.K.}
\affiliation{$^2$AstroParticule \& Cosmologie,
UMR 7164-CNRS, Universit\'e Denis Diderot-Paris 7,
CEA, Observatoire de Paris,
10 rue Alice Domon et L\'eonie
Duquet, F-75205 Paris Cedex 13, France}
\affiliation{$^3$
Department of Physics, National Dong Hwa University, Hua-Lien,
Taiwan 974, R.O.C.  }
\affiliation{$^4$ Department of Physics and Helsinki Institute of Physics, PL 64, FI-00014 University of Helsinki, Finland}

\date{\today}
\begin{abstract}
The BEC regime of a cold fermi gas is characterised by coupled atoms (dimers) which, superficially, look like elementary bosons. We examine how simply-bosonic they really are; firstly, in the Bogoliubov approximation and further, through new  actions for the BEC regime in which dimers are represented by coupled Gross-Pitaevskii fields. We find identity at the level of the Bogoliubov approximation in the deep BEC regime, permitting a simple Gross-Pitaevskii description. This fails rapidly as we move towards the BCS regime. However, even in the deep BEC regime there is an intrinsic difference if we go beyond the Bogoliubov approximation.  To exemplify this we construct vortex solutions.
\end{abstract}


\maketitle

\section{Introduction}

Condensates of ultra-cold fermi gases show a range of behaviour, interpolating from a BCS regime characterised by Cooper pairs to a BEC regime characterised by diatoms or dimers~\cite{fermigas}.  In this paper we examine the difference between these di-fermi condensates and those of {\it elementary} (point-particle) bosons, particularly in the BEC regime.

We do this by contrasting the fermi gas partition function to the bosonic Gross-Pitaevskii (GP) functional integral, whose representation of bosonic condensates is wholly familiar and whose equivalence to the path integral partition function of elementary bosons is well understood \cite{wiegel}. We have already seen \cite{lee3} that, with some qualifications, condensates of cold fermi gases are described by a Gross-Pitaevskii model of simple bosons in the acoustic (or hydrodynamic) approximation of phonons with linear dispersion relation $\omega = ck$.  The first result of this paper is to show that, in the deep BEC regime, the representation of dimer condensates by condensates of simple bosons is further valid in the Bogoliubov picture (of bogolons) with dispersion $\omega^2 = c^2k^2 + (k^2/2M)^2$, for which phonons provide the long wavelength limit and dimers the short wavelength limit.

Superficially, this looks a surprising result in that the degrees of freedom of elementary bosons and di-fermions are different; the former permits only a gapless mode (the bogolon, labelled $\theta$) whereas the latter has, in addition, a gapped mode (the dimensionless dimer density fluctuation, labelled $\epsilon$). In the BEC regime, and there alone, the gapped mode becomes an auxiliary field with no degrees of freedom and a comparison of dimer to elementary boson can be made. We do more, identifying the bogolon $\theta$ as the  non-relativistic quantum excitation associated with the BEC (sound-cone) relativistic fluctuation fields (essentially $\theta \pm i\epsilon$) of the condensate \cite{evans, rivers1}. 
It is very clear that  there is no simple relationship between di-fermions and simple bosons away from the deep BEC regime.

 However, the main result  of this paper is to provide a generalised Gross-Pitaevskii description of fermi gases beyond the Bogoliubov picture. To do so we adopt the approach of Aitchison, Thouless {\it et al.} \cite{aitchison} for fermi gases in the deep BCS regime of Cooper pairs (based on the fields $\theta\pm\epsilon$), to which our more general fermi gas, controlled through an explicit Feshbach resonance, provides a controllable extension to the BEC regime.  Di-fermions are now represented by a pair of coupled Gross-Pitaevskii fields, to which the deep BEC regime provides great simplification. We now see a difference between elementary bosons and dimers in the BEC regime even when density fluctuations are auxiliary.  We stress that these go beyond Gaussian fluctuations about mean-field results. Nonetheless, to understand this difference better we shall look for vortex solutions, to compare with conventional Gross-Pitaevskii vortices (e.g.\cite{fetter, kleinert, rorai}) within the mean-field framework.
In practice, despite the universally different equations that vortices of elementary bosons and of fermi gas BEC composite bosons satisfy we find near-identity of the solutions in the deep BEC regime.  In all that follows we only work at {\it zero} temperature in natural units with $\hbar=1$.
\\
\\
Because of the intricacy of the subsequent discussion we offer the following brief guide:

\noindent The natural framework in which to make comparisons is that of functional integral partition functions realised through Lagrangians. Although the argument is straightforward the notation can become cumbersome with so many intermediate steps. To keep it as simple as possible we adopt the following consistent notation. Apart from the point particle partition function, each of the several partition functions that we shall introduce carries a label, usually as a subscript (e.g. B for Bogoliubov), which characterises the approximation or reformulation that we are making. For a label X, say, we write the partition function
\be
Z_X = \int {\cal D}[{\rm fields}]\,e^{iS_X[{\rm fields}]}
\ee
for the fields corresponding to the relevant degrees of freedom. When $S_X$ is {\it local}, we can define  the Lagrangian density $L_X(\rm{fields (x)})$ by
\be
S_X[{\rm fields}] = \int d^4x\, L_X({\rm fields (x)})
\ee
in which $d^4x \equiv dt~d^3x$.  Since we are continually moving between partition functions, actions and Lagrangians this leads to an economical use of notation.

Finally, we stress the importance of preserving the underlying symmetry of the theory, in this case the Galilean group. In particle physics, where the Lorentz group is the invariance group, we would not make approximations for relativistic systems that violate Lorentz symmetry. Equally, we preserve Galilean invariance step by step.

\section{Simple bosons revisited}

In order to contrast dimers to bosons, it is will be useful to first review elementary bosons. We do this in some detail so as to make later comparisons transparent.
Consider elementary bosons, of mass $M_b$, with two-body potentials $V(|{\bf x|})$, for which the $n$-boson Hamiltonian is ($i, j = 1,2,3, ...,n$)
\be
H_n ({\bf p_i, x_i}) = \sum_i\frac{{\bf p}_i^2}{2M_b} + \sum_{i> j}V(|{\bf x}_i - {\bf x}_j|).
\label{HB}
\ee
The question is to what extent can a condensate of dimers be represented by a Hamiltonian of the form (\ref{HB})?
We rephrase the question by exploiting the duality between the bosonic Feynman {\it path} integrals  for ensembles of bosons described by (\ref{HB}) and the {\it functional} integral for the Gross-Pitaevskii quantum field \cite{wiegel}. On taking the trace over n-particle bosonic states of (\ref{HB}) the zero-temperature ($\beta \rightarrow \infty$) limit of the grand canonical partition function  (with subscript $b$ for {\it boson})
\be
 Z_b = {\rm Tr}\exp{-\beta(\hat{H}_n - \mu_0 \hat{n})}
 \label{Zb}
 \ee
can be written without approximation as the quantum field functional integral \cite{wiegel}
\begin{equation}
  Z_b \equiv Z_{GP} = \int {\cal D}\psi{\cal D}\psi^*
  \,e^{i S_{GP}[\psi^*,\psi]},
  \label{Z}
 \end{equation}
where
 \begin{equation}
S_{GP}[\psi^*,\psi] = \int d^4x \,L_{GP}(\psi^*,\psi) \equiv
\int d^4x \,\bigg\{i\psi^{\ast}\dot{\psi} - \frac{1}{2M_b}\nabla\psi^{*}
\cdot\nabla\psi - \frac{1}{2N_b}(\psi^*\psi - \rho_0^b)^2 \bigg\}
\label{SGP}
\end{equation}
is the Gross-Pitaevskii action for the order-parameter field $\psi$. In (\ref{SGP}) $\rho_0^b$ is the boson density and we have assumed a point potential $V$ of strength $N_b^{-1}$ \cite{wiegel}.
Thus, if we can recast our model of dimers in the form (\ref{Z}) and (\ref{SGP}) they have a realisation as simple bosons.

To see that the (only) quanta of such elementary particle condensates are gapless `bogolons', identifiable as phonons in the long wavelength/low momentum limit, we use the Madelung decomposition
$\psi = \sqrt{\rho} e^{i\theta}$.
 $L_{GP}(\psi^*,\psi)$ then becomes, in terms of the Galilean invariants $\rho$ and $G(\theta) \equiv {\dot\theta} + (\nabla\theta)^2/2M_b$,
\be
L_{GP}(\theta, \rho) = -\rho G(\theta)
-\frac{1}{8M_b\rho}(\nabla\rho)^{2} - \frac{1}{2N_b}(\rho - \rho_0^b)^2.
\label{LGP}
\ee
The familiar acoustic (or hydrodynamic) approximation for long wavelength phonons corresponds to dropping the density fluctuation second term in (\ref{LGP}). Here, however, we adopt the less restrictive Bogoliubov approximation (with superscript $B$ for {\it Bogoliubov}) of retaining all quadratic fluctuations in the density,
\be
L^{B}_{GP}(\theta, \rho) = -\rho G(\theta)
-\frac{1}{8M_b\rho_0^b}(\nabla\rho)^{2} - \frac{1}{2N_b}(\rho - \rho_0^b)^2
\label{L0GP}
\ee
in which, in the 2nd term, the $\rho$ in the denominator has been replaced by its mean-field value $\rho_0^b$.
With unit Jacobian the partition function $Z_b$ of (\ref{Z}) is approximated by
\begin{equation}
  Z^B_b \equiv Z^B_{GP} =  \int {\cal D}\theta {\cal D}\epsilon\,e^{i S^B_{GP}[\theta, \epsilon]}.
  \label{ZBthetaeps}
 \end{equation}
  Since $L^{B}_{GP}$ is quadratic in $\rho$ we perform the $\rho$ integration to give
 \begin{equation}
  Z^B_b =  \int {\cal D}\theta\,e^{i S^B_b[\theta]}
  \label{ZBtheta}
 \end{equation}
 where $S^B_b[\theta]$ is the familiar (e.g. see \cite{lancaster}) {\it non-local} action
 \be
 S^B_b[\theta] = \int d^4 x ~(-\rho_0^b G(\theta (x))) + \int\int d^4 x~d^4 y \frac{N_b}{2} G(\theta (x))\Delta^0 (x-y) G(\theta (y)).
 \label{SB}
 \ee
Here $\Delta^0 (x)$ is the {\it instantaneous} density-fluctuation Yukawa potential
 \be
 \Delta^0 (x) = \frac{1}{(2\pi)^4}\int d\omega~d^3 k~\frac{e^{i\omega t}e^{-i\bf k.\bf x}}{1 + N_b {\bf k}^2/4\rho_0^b M_b}
 \label{D0}
 \ee
which mediates phonon-phonon scattering via the $(\nabla\theta)^2$ term in $G$.

 To describe the quasi-particle spectrum we only need the quadratic part of $S^B_b[\theta]$,
 \be
 S^{B(2)}_b[\theta] = \int d^4 x ~\bigg(-\frac{\rho_0^b}{2M_b}(\nabla\theta)^2\bigg) + \frac{N_b}{2} \int\int d^4 x~d^4 y\, {\dot\theta} (x)\Delta^0 (x-y) {\dot\theta} (y).
 \label{SBQ}
 \ee
Plane wave solutions give the dispersion relation in $(\omega, {\bf k})$ as
\be
\frac{\rho_0^b {\bf k}^2}{M_b} = \frac{N_b\omega^2}{1 + N_b {\bf k}^2/4\rho_0^b M_b},
\label{dis}
\ee
which can be rewritten in the more familiar Bogoliubov form
\be
\omega^2 = c_b^2 {\bf k}^2 + \bigg(\frac{{\bf k}^2}{2M_b}\bigg)^2,
\label{dis2}
\ee
where $c_b^2 = \rho_0^b/M_bN_b$ is the phonon speed of sound.
 The more restrictive {\it ultra-local} acoustic/hydrodynamic action, which will play a role later, of ignoring gradient terms $(\nabla\rho)^2$ in (\ref{L0GP}) is
  \ba
 S^{\rm ac}_b[\theta] &=& \int d^4 x ~(-\rho_0^b G(\theta (x))) + \int\int d^4 x~d^4 y \frac{N_b}{2} G(\theta (x))\delta^4 (x-y) G(\theta (y))
 \nonumber
 \\
  &=& \int d^4 x ~(-\rho_0^b G(\theta (x))) + (N_b/2) G(\theta (x))^2)
 \label{Sac}
 \ea
corresponding to replacing $\Delta^0(x)\to \delta^{4}(x)$, with dispersion relation $\omega^2 = c_b^2 {\bf k}^2$.

  Finally, for comparison, it is useful to rewrite
  (\ref{SB})  in terms of $c_b$ as
  \be
 S^{B(2)}_b[\theta] = \int d^4 x ~(-\rho_0^b G(\theta(x))) +  \frac{\rho_0^b}{2M_bc_b^2} \int\int d^4 x~d^4 y\, G(\theta(x))\Delta^0 (x-y) G(\theta(y)),
 \label{SB2}
 \ee
 with, from (\ref{D0})
 \be
 \Delta^0 (x) = \frac{1}{(2\pi)^4}\int \frac{d\omega~d^3 k~e^{i\omega t}e^{-i\bf k.\bf x}}{1 + {\bf k}^2/4 M_b^2c_b^2} =
 \frac{(2M_bc_b)^2\delta(t)}{4\pi|{\bf x}|}e^{-2M_bc_b|{\bf x}|}
 \label{D1}
 \ee
 and, going back a step,
 \begin{equation}
S_{GP}[\psi^*,\psi] = \int d^4x \,L_{GP}(\psi^*,\psi) =
\int d^4x \,\bigg\{i\psi^{\ast}\dot{\psi} - \frac{1}{2M_b}\nabla\psi^{*}
\cdot\nabla\psi - \frac{M_bc_b^2}{2\rho_0^b}(\psi^*\psi - \rho_0^b)^2 \bigg\}.
\label{SGPc}
\end{equation}

In the next section we show how a cold fermi gas condensate in the deep BEC regime, under an analogous Bogoliubov approximation,  reproduces (\ref{SB}) and (\ref{SGPc}) and hence can be construed as a condensate of elementary dimers in that approximation.

\section{The Cold Fermi Gas Model}

 There is a huge literature on tunable cold fermi gases but, for analytical simplicity, we consider only fermi atoms $\psi_{\sigma}$ with spin label $\sigma=(\uparrow , \downarrow)$, mass $m_f$ and chemical potential $\mu$, interacting through an idealised {\it narrow} s-wave Feshbach resonance. This restricts the literature (e.g. \cite{gurarieplus,gurarie}) beyond our own work \cite{lee3,lee4,lee5}, to which we shall refer. To be concrete, we have in mind the resonance in $^6\rm{Li}$ at $H_0 = 543.25 G$, discussed in some detail in \cite{strecker}.

All calculations are performed at temperature $T=0$ where, in the absence of external fields, the partition function for di-fermions
\begin{equation}
  Z_{2f} = \int {\cal D}\psi{\cal D}\psi^*{\cal D}\varphi{\cal D}\varphi^*
  \,e^{i S_{2f}[\psi^*,\psi,\varphi^*,\varphi]}
  \label{ZFF}
 \end{equation}
is expressed through the action \cite{gurarie}
\begin{eqnarray}
S_{2f}[\psi^*,\psi,\varphi^*,\varphi] &=& \int dt\,d^3x\bigg\{\sum_{\uparrow , \downarrow}
 \psi^*_{\sigma} (x)\ \left[ i \
\partial_t + \frac{\nabla^2}{2m_f} + \mu \right] \ \psi_{\sigma} (x)
\nonumber \\
   &+& \varphi^{*}(x) \ \left[ i  \ \partial_t + \frac{\nabla^2}{2M_D} + 2 \mu -
\nu \right] \ \varphi(x) \nonumber \\
&+& g \left[ \varphi^{*}(x) \ \psi_{\downarrow} (x) \ \psi_{\uparrow}
(x) + \varphi(x)  \psi^{*}_{\uparrow} (x) \ \psi^{*}_{\downarrow} (x)
\right]\bigg\}. \label{Lin}
\end{eqnarray}
 The Feshbach resonance, with binding energy $\nu$, is described by $\varphi(x)$ and its relation to dimers is already present in the mass of the dimer in the deep BEC regime, $M_D =2m_f$. It is not difficult to accommodate point s-wave interactions \cite{lee1,lee2} in (\ref{Lin}) but it complicates the formalism hugely even though it leads to some unchanged outcomes e.g. the behaviour of the speed of sound in the deep BEC and BCS regimes.
 We have examined the model of \cite{gurarie} in some detail in previous papers \cite{lee3,lee4,lee5}, from which we borrow where appropriate.

Since the action is quadratic in the Fermi fields, they can be integrated out to give the exact (one fermi-loop) Feshbach (F) partition function
\begin{equation}
  Z_{2f}\equiv Z_{F} = \int {\cal D}\varphi{\cal D}\varphi^*
  \,e^{i S_F^{NL}[\varphi^*,\varphi]}
  \label{ZF}
 \end{equation}
with {\it non-local} (NL) action
  \begin{eqnarray}
 S_F^{NL}[\varphi^*,\varphi] &=& -i\,{\rm Tr}\ln {\cal G}^{-1}[\varphi^*,\varphi]
 +\int dt~d^3x\,\varphi^{*}(x) \ \left[ i  \ \partial_t + \frac{\nabla^2}{2M_D} + 2 \mu - \nu\right]\varphi(x),
\label{SNL}
 \end{eqnarray}
in which
  ${\cal G}^{-1}$ is the fermi loop
inverse Nambu Green function,
\begin{eqnarray}
 {\cal G}^{-1} &=& \left( \begin{array}{cc}
        i \partial_t - \varepsilon         &g\varphi(x) \\
                  g\varphi^{*}(x) & i \partial_t +
                  \varepsilon
                  \end{array} \right).
                  \end{eqnarray}
Here $\varepsilon = - \nabla^2/2m_f - \mu $ (or in momentum space, $\varepsilon_k = {\bf k}^2/2m_f - \mu$)  is the energy measured from the Fermi surface.

The action (\ref{SNL}) is invariant under global $U(1)$ transformations $\theta \rightarrow \theta+c$ for some constant $c$, but the variational equation $\delta S^{NL}_F/\delta\varphi^*(x) = 0$ permits constant solutions $\varphi (x) = \varphi_0$ which break this symmetry spontaneously, satisfying
\be
\int \frac{d^3{\bf p}}{(2\pi)^3}
\bigg( \frac{1}{2E_p} - \frac{1}{2\epsilon_{p}} \bigg) = \frac{\nu - 2\mu}{g^2}
\equiv -\frac{m_f}{4\pi a_S}
\label{gap}
\ee
after additive UV renormalisation of $\nu$, where $E_{p}=(\varepsilon_{p}^2 + g^2|\varphi_{0}|^2 )^{1/2} $ and $\epsilon_{p} = {\bf p}^2/2m_f$.
The second equation in (\ref{gap}) defines the atomic s-wave scattering length $a_S$.  The advantage of narrow Feshbach resonances is that $\nu$ and $\mu$, and hence $1/k_Fa_S $, are tunable with an external magnetic field $H$ approximately as
$1/k_Fa_S \propto (H - H_0)$
where $H_0$ determines the unitary limit of infinite scattering length \cite{strecker}. The BEC and BCS regimes correspond to $1/k_Fa_S > 0$ and $1/k_Fa_S < 0$ respectively, and the deep BEC regime in which we are interested corresponds to large positive $1/k_Fa_S$.
Finally, the fermion number density $\rho_0^f$ is
\begin{equation}
 \rho_0^f = \rho^{\rm{ex}}_0 +
2\rho^D_0 , \label{f-density}
\end{equation}
  where
$ \rho^{\rm{ex}}_0 = \int d^3 {\bf p} / (2\pi)^3 \ \left[ 1 -
\varepsilon_{p}/E_{p}
  \right] $
is the explicit fermion density and $\rho^D_0 = |\varphi_{0}|^2$. The suffix $D$ reminds us that in the deep BEC regime, where $\rho^{ex}_0$ vanishes, $\rho^D_0$ is the dimer density (two fermions per dimer).

The condensate of the theory is
\be
\,\varphi (x) = |\varphi(x)| \ e^{i
\theta (x)}
\label{modphi}
\ee
with the gapless mode encoded in the phase $\theta$.
The Galilean invariants of the theory are
 $|\varphi|$ itself, $G(\theta) = \dot{\theta} + (\nabla
\theta )^2/2M_D$ as before, and the comoving time derivative of $|\varphi|$ in a fluid with fluid velocity $\nabla \theta/M_D$,
 \be
D_t(|\varphi|,\theta ) =
\partial_t|\varphi|+\nabla \theta .\nabla |\varphi|/M_D.
\ee
On changing variables to $\theta$ and $|\varphi|$, so that
$S_F^{NL}[\varphi,\varphi^{*}]\equiv S_F^{NL}[\theta,|\varphi|]$ the partition function becomes
\be
Z_{2f}\equiv Z_F = \int {\cal D}\theta{\cal D}|\varphi|\,
e^{iS_F^{NL}[\theta,|\varphi|]}
  \,
  \label{Zd}
\ee
where $S_F^{NL}[\theta,|\varphi|]$ incorporates the Jacobian functional determinant $J[|\varphi|] = {\rm Det}[|\varphi|] = \exp\{Tr\ln|\varphi|\}$ due to the change of variables.

We now see that preserving Galilean invariance step by step leads to an interesting problem. The gap equation (\ref{gap}) followed from the variational equation $\delta S^{NL}_F[\varphi,\varphi^{*}]/\delta\varphi^* = 0$, which is not Galilean covariant.
However, the Galilean covariant equation $\delta S_F^{NL}[\theta,|\varphi|]/\delta|\varphi|=0$ incorporating the Jacobian gives the same solution (\ref{gap}) in terms of the physically observable $\mu,\nu$ {\it after} further UV renormalisation of $\nu$.

We now make a Galilean-invariant decomposition
\be
\varphi(x) = (|\varphi_0| + \delta |\varphi(x)|)e^{i\theta(x)}
\label{varphi}
\ee
where $\delta |\varphi| = |\varphi|  -  |\varphi_{0}| = \delta\rho_0^D/4|\varphi_0|$ is a {\it small}
fluctuation in the reduced condensate density, whereas $\theta (x)$ is not small.
On changing variables to $\theta$ and and the dimensionless $\epsilon = \kappa^{-1/2}\delta|\varphi|$,
we expand the resulting $S_F^{NL}[\theta,\epsilon] (\equiv S_F^{NL}[\theta,|\varphi|]/\delta|\varphi|]$ in powers of the Galilean invariants along the lines taken in \cite{aitchison} and sketched in \cite{lee1,lee4,lee5}.
Retaining all squares of Galilean invariants all the terms relevant to the immediate discussion are contained in the {\it local} effective action $S_F^{\rm eff}[\theta, \epsilon]$,
      \begin{eqnarray}
S_F^{\rm eff}[\theta, \epsilon] = \int d^4x L_F^{\rm{eff}}(\theta, \epsilon) = \int d^4x\bigg[ \frac{N_f}{4}\ G^2(\theta, {\epsilon}) -\frac{1}{2}{\rho^f_0}
   G(\theta, {\epsilon})
  -{\alpha}{\epsilon}G(\theta, {\epsilon})
 +\frac{1}{4}{\eta}D_t^2({\epsilon},\theta)
  -\frac{1}{4}{\bar M}_0^2{\epsilon}^2\bigg].
 \label{Leff}
 \end{eqnarray}
 Here the scale factor $\kappa$
has been chosen so that on extending
$G(\theta)$ to $G(\theta, \epsilon ) = \dot\theta + (\nabla
\theta )^2/2M_D + (\nabla \epsilon )^2/2M_D$ in (\ref{Leff}), $\epsilon$ has the
 same coefficients as $\theta$ in its spatial derivatives \cite{aitchison} and all coefficients are defined in terms of physical variables.
The behaviour of $\kappa$ together with that of the coefficients in (\ref{Leff}), namely $\eta,\bar{M}_0, \alpha, \rho_0^D, N_f$, are given in the Appendix. In particular, the quadratic terms in $\epsilon$ due to the Jacobian are eliminated by the same redefinition of $\nu$ adopted earlier and, until explicitly stated, we shall ignore the Jacobian henceforth. A similar ability to ignore the Jacobian  is seen in renormalisable relativistic field theory \cite{salomonson}.

The action $S_F^{\rm eff}[\theta, \epsilon]$ contains (derivative) terms in $\epsilon$ up to fourth order. To bring our model into correspondence with the Bogoliubov approximation of the previous section we retain only quadratic terms in its derivatives, with action $S_F^B[\theta, \epsilon]$, obtained simply from (\ref{Leff}).  Integrating over $\epsilon$ in the resulting Bogoliubov-equivalent partition function
 \be
Z_F^{B}= \int {\cal D}\theta{\cal D}\epsilon\,e^{iS_F^{B}[\theta,\epsilon]} = \int {\cal D}\theta\,
e^{iS_F^{B}[\theta]}
  \,
  \label{ZB0}
\ee
gives the {\it non-local} action
\begin{eqnarray}
&& S^B_{F}[\theta] = \int d^4x \left\{-\frac{1}{2}{\rho^f_0}
   G(\theta (x) )    \right\}
 +\int d^4x \int d^4x' G(\theta
 (x))\,\bigg\{\frac{N_f}{4}\,\delta^4(x - x') + \frac{\alpha^2}{{\bar M}_0^2} \Delta_F (x-x')\bigg\} \, G(\theta
 (x')) \, , \nonumber\\
 \label{SB2}
 \end{eqnarray}
 where $\Delta_F(x)$ ($F$ for Feshbach) is
\begin{eqnarray}
\Delta_F (x) = \frac{1}{(2\pi)^4}\int d\omega~d^3k\frac{e^{i\omega t}e^{-i{\bf k.\bf x}}}{1 -\eta\omega^2/{\bar M}_0^2 + \rho_0^f k^2/M_D + {\bar M}_0^2  }.
\nonumber
\\
\label{DF}
\end{eqnarray}
The normalisation of $\Delta_F(x)$ is chosen so that, in the acoustic approximation, in which we set $\omega = k =0$ in its integrand, $\Delta_F(x) = \delta^{4}(x)$.
In comparison to the Bogoliubov approximation of (\ref{SB}), (\ref{SB2}) has a contact term contributing directly to phonon-phonon scattering and a non-instantaneous density fluctuation propagator.

Retaining only the quadratic field terms gives
\begin{eqnarray}
&& S^{B(2)}_{F}[\theta] = \int d^4x \left\{ -\frac{\rho_0^f}{4M_D}(\nabla\theta(x)^2)  \right\}
 +\int d^4x \int d^4x' \dot\theta
 (x)\,\bigg\{\frac{N_f}{4}\,\delta^4(x - x') + \frac{\alpha^2}{{\bar M}_0^2} \Delta_F (x-x') \bigg\}\, \dot\theta
 (x') \, . \nonumber\\
 \label{SB22}
 \end{eqnarray}
From (\ref{SB22}) we see that the quasi-particle spectrum now satisfies
\be
 \vspace{1cm}
 \frac{\rho_0^f}{M_D}{\bf k}^2 = \omega^2\bigg(N_f + \frac{4\alpha^2}{-\eta\omega^2 + \rho_0^f{\bf k}^2/M_D + {\bar M}_0^2}\bigg).
 \label{dis3}
 \ee
It follows directly that the speed of sound $c$ for long wavelength modes is given as
\be
c^2 \equiv \frac{\rho_0^f}{M_D N} = \frac{\rho_0^f/M_D}{N_f + \frac{4\alpha^2}{\bar M_0^2}}
\label{c2}
\ee
where $N = N_f + 4\alpha^2/\bar M_0^2$. As it stands, for generic $1/k_Fa_S$ (\ref{dis3}) can be expanded as
\be
\omega^2 = c^2 k^2 + c_1^2 k^4 + \ldots
\label{dispersion}
\ee
where $c^2$ is as in (\ref{c2}) and
\ba
c_1^2 &=&\frac{ ({\rho_0^f/M_D})^2 }{ \left(N_f + \frac{4\alpha^2}{\bar{M}^2}\right)^3} \left( \frac{4\alpha^2}{\bar{M}^4} \right) \left(N_f - \eta + \frac{4\alpha^2}{\bar{M}^2} \right).
\label{c1}
\ea
A priori this does not resemble the Bogoliubov dispersion relation and the action (\ref{SB2}) does not resemble its simple bosonic counterpart (\ref{SB}). However, we shall see later that in the the deep BEC regime they do become identical.

The sound speed $c^2$ is  to be contrasted to what we term the mean-field speed of sound
\be
c_0^2 = \rho_0^f/M_DN_f,
\label{C02}
\ee
obtained by ignoring density ($\epsilon$) fluctuations (i.e. taking $\alpha = 0$ in the above, which we shall see plays a dominant role in subsequent calculations.

\subsection{The deep BEC regime}

In the Appendix we show characteristic behaviour of the coefficients and $c^2$ across the whole BCS/BEC regime as a function of $1/a_Sk_F$. The key attributes of the coefficients relevant to the deep BEC are that (see Figs. 2) as $1/a_S k_F\rightarrow\infty$:
\\
\\
i)  the mode-coupling coefficient $\alpha\rightarrow \rho_0^f\to 2\rho_0^D$.
\\
\\
ii) both $N_f$ and $\eta \approx N_f\rightarrow 0$, with $N_f/\eta\rightarrow 1$.
\\
\\
iii) ${\bar M_0}^2\rightarrow 0$ and
\\
\\
iv) $\kappa^{-1}|\varphi_0|^2\rightarrow 1$.
\\
\\
We note the very different behaviours of $N_f$ and $N$ in the deep BEC regime. Whereas $N_f$ vanishes there $N$ does not.
In particular, in the idealised model (\ref{Zd}) the vanishing of ${\bar M}_0^2$ in the BEC limit forces the true speed of sound $c\rightarrow 0$ because of the {\it divergence} of $N$ whereas the mean-field speed of sound $c_0\rightarrow\infty$ itself {\it diverges}.

In reality the speed of sound does not vanish for the accessible BEC regime~\cite{joseph}.
To induce non-vanishing $c$ we include a term $L_D^{\rm eff} = -u_B |\varphi (x)|^4/4$ in the integrand of (\ref{Lin}) to incorporate direct dimer-dimer interactions \cite{holland,timmermans}.
The gap equation for the combined action ${\bar S}_F^{NL}[\theta,|\varphi|] = S_F^{NL}[\theta,|\varphi|] + S_D^{\rm eff}[|\varphi|]$ now becomes
\be
\int \frac{d^3{\bf p}}{(2\pi)^3}
\bigg( \frac{1}{2E_p} - \frac{1}{2\epsilon_{p}} \bigg) + \frac{u_B}{2g^2}|\varphi_0|^3 = \frac{\nu - 2\mu}{g^2}
\equiv -\frac{m_f}{4\pi a_S}
\label{gap2}
\ee
The Appendix shows the dependence of the parameters of the model on $u_B$. In particular we see that, in the deep BEC regime, properties i) - iv) above are unaltered.
However, the inclusion of $L_D^{\rm eff}$ changes the coefficient ${\bar M}^2$ of the quadratic term in $\epsilon$ in (\ref{Leff}) as ${\bar M}_0^2\rightarrow {\bar M}^2 = {\bar M}_0^2 + 6 u_B\kappa|\varphi_0|^2$ whence, in the deep BEC regime $1/k_Fa_S\to\infty$,
 \be
 {\bar M}^2\to  {\bar M}_{\infty}^2 = 6 u_B|\varphi_0|^4.
\label{barM}
\ee
on using iv).

As our final step in establishing dimer/boson identity we return to (\ref{SB2}), in which we now replace  $c^2\to c_{\infty}^2 = {\bar M_{\infty}}^2/4M\rho^f_0 \neq 0$ to contrast it to its simple bosonic counterpart (\ref{SB}). In the BEC limit with $N_f\approx \eta\to 0$, the acoustic $\delta$-function in (\ref{SB2}) and (\ref{SB22}) vanishes and $\Delta_F(x)$ becomes the Yukawa $\Delta^0(x)$ of (\ref{D0}). $S^B_{F}[\theta]$ now takes the limit
\begin{eqnarray}
&& S^B_{F}[\theta] \to \int d^4x ( -{\rho}^D_0
   G(\theta (x) )
 +\frac{\rho^D_0}{2M_Dc_{\infty}^2}\int d^4x \int d^4x' G(\theta
 (x))\, \Delta^0 (x-x') \, G(\theta
 (x')) \, ,
 \label{SB3}
 \end{eqnarray}
{\it identical} to $S_{b}^{B}[\theta]$ of  (\ref{SB2}) on substituting dimers for elementary bosons.
In particular, the gapless dispersion relation (\ref{w-}) takes Bogoliubov limiting form
\begin{eqnarray}
\omega^2 \to c_{\infty}^2k^2 + (k^2/2M_D)^2.
\end{eqnarray}
This is our first result, that the partition function $Z_F^B$ obtained from the cold fermi gas action $Z_{2f}$ of(\ref{Lin}) in the deep BEC regime has the form of the partition function $Z_b^B$ of (\ref{ZBtheta}) for elementary bosons
in the Bogoliubov approximation for which the quasi-particle dispersion relation is (\ref{dis2}) and the action (\ref{SB}) (or (\ref{SB2}).

As we move away from the deep BEC regime the difermions are less and less like elementary bosons and their equations of motion less and less like the GP equation. As $N_f$ and $\eta$ cease to be small the acoustic term in (\ref{SB2}) and (\ref{SB22}) increases while, at the same time, $\Delta_F(x)$ ceases to be instantaneous. See the Appendix for details.

However, there is simplification in the deep BSC regime $1/k_Fa_S\to -\infty$ if we extend the analysis of the idealised model there where $\alpha\to 0$, ${\bar M}_0^2\neq 0$ and $N_f = \eta$ is the density of states at the Fermi level \cite{aitchison}. The resulting speed of sound $c_{-\infty}^2 = \rho_0^f/M_DN_f = v_F^2/3$ \cite{aitchison}. In that case, with the coefficient of $\Delta_F$ vanishing in (\ref{SB2}) and (\ref{SB22}), we can only recover the {\it acoustic}  approximation of (\ref{Sac}), with linear dispersion relation (as follows directly from (\ref{c2})), which does {\it not} describe elementary bosons. However, we now understand why the acoustic approximation is valid across the whole BCS-BEC range in this model if that is sufficient.

\section{Gross-Pitaevskii equations for $\theta + i\epsilon$ }

 Before attempting to go beyond the Bogoliubov picture we can do more to identify the dimer field better, in particular with regard to the fluctuation (Feshbach) field $\delta\varphi = \varphi - \varphi_0$, to which it is expected to be related when dimers dominate, despite the fact that $\delta\varphi$ has no simple behaviour under Galilean transformations. This relation is necessarily complicated because the dynamics of the theory is carried in the field phase and not its fluctuations \cite{aitchison}. Nonetheless, it is worth showing the relationship since it is not uncommon (e.g. \cite{devreese,stoof}) to see derivative expansions in fluctuation fields despite Galilean invariance being crucial to some attributes of the quantum behaviour (e.g. the stochastic nature of the sound-cone metric \cite{lee6}).

 We have already commented upon the very different behaviours of $N_f$ and $N$ in the deep BEC regime. Whereas $N_f$ vanishes there $N$ does not. The effect is very simple in the case $u_B = 0$ when $c\to 0$ in the deep BEC regime and we consider that first.
Then, because of the vanishing of $N_f$ and $N_f = \eta$ there, (\ref{dis3}) can be written as
\be
\omega = \sqrt{{\cal M}^2c_0^4 + k^2c_0^2} - {\cal M}c_0^2 \rightarrow k^2/2M_D
\label{w-}
\ee
where ${\cal M} = (\alpha/\rho_0^f)M_D\rightarrow M_D$ as $c_0\rightarrow\infty$. We stress that (\ref{w-}) is not a high momentum limit but a deep BEC high-$c_0$ limit, recreating the (Bogoliubov) free-particle spectrum in the $c\rightarrow 0$ limit.

Although the GP model is that of a complex field, because of Galilean invariance it describes only a single mode, the (gapless) phonon/bogolon. This is not the case for our fermi system (\ref{Zd}), which describes a Higgs-Goldstone model with a gapped mode, related to $\epsilon$, identified with quantised density fluctuations.
 The dispersion relation of the gapped mode is most simply derived by retaining only the quadratic terms in (\ref{Leff}) and integrating over $\theta$. The outcome is that, as we approach the deep BEC regime, then
\be
\omega = \sqrt{{\cal M}^2c_0^4 + k^2c_0^2} + {\cal M}c_0^2 \rightarrow 2M_Dc_0^2\rightarrow\infty
\label{w+}
\ee
as $c_0 \to\infty$. That is, the gapped mode decouples in the deep BEC limit, leaving us with a single (gapless) phonon mode akin to that of the simple GP model, as we have seen.

The relations (\ref{w-}) and (\ref{w+}) suggest that the deep BEC regime is explicable through a (sound-cone) relativistic theory of particles of mass $M_D$, rest mass energy $M_Dc_0^2$ in a critical chemical potential $M_Dc_0^2$ in the limit $c_0\rightarrow\infty$  \cite{rivers1,evans}. The situation is qualified slightly if $c\to c_{\infty}\neq 0$, but it is not difficult to identify the field whose quanta are the bogolons.

All the terms relevant to particle identification are contained in the (non-invariant) quadratic effective
action obtained from the action (\ref{Leff}) \cite{lee3,lee4,lee5} (or equivalently, $S_F^B[\theta, \epsilon]$)
\begin{eqnarray}
 S_{F}^{B(2)}[\theta, \epsilon] &=& \int d^4x \frac{1}{4}\bigg\{N_f\dot{\theta}^2 -\frac{\rho^f_0}{M_D}(\nabla\theta)^2 - 2\alpha(\epsilon\dot\theta - {\dot\epsilon}\theta) + \eta_0\dot{\epsilon}^2
-\frac{\rho^f_0}{M_D}(\nabla\epsilon)^2 - {\bar M}^2{\epsilon}^2\bigg\}.
 \label{LeffU0}
 \end{eqnarray}
 The kinetic part of the action (\ref{LeffU0}) looks like a (sound-cone) relativistic theory  for a complex field
\be
\phi(x) = \frac{1}{\sqrt{2}}(\theta (x) + i\epsilon (x))
\label{phi}
\ee
in the presence of a chemical potential if we write the Lagrangian density from (\ref{LeffU0}) as
\be
L^{(2)}\equiv L_F^{B(2)}[\phi,{\phi}^\dagger] = \frac{1}{2}\bigg\{N_f \left[ \dot{\phi}^\dagger \dot{\phi} - c_0^2 \nabla \phi^\dagger \nabla \phi \right] + i \alpha(\dot{\phi}^\dagger \phi  - \dot{\phi} \phi^\dagger )   + \frac{1}{4} \bar{M}^2(\phi - \phi^\dagger)^2\bigg\}
\label{newL}
\ee
 where we have taken $N_f = \eta$, appropriate for the deep BEC regime (and dropped further labels on $L^{(2)}$). [The results are unchanged if we take $N_f = \eta[1 +\Delta]$  as $N_f,\eta,\Delta\to 0$.] The speed of propagation in (\ref{newL}) is the diverging $c_0$. We note that when ${\bar M}^2 = {\bar M}_0^2\rightarrow 0$ the Lagrangian
(\ref{newL}) has a global $U(1)$ invariance associated with the conservation of
 dimer number.

From (\ref{newL}) the canonical  momenta are given by
\ba
\pi &=& \frac{\partial L^{(2)}}{\partial \dot{\phi}^\dagger} = \frac{1}{2}\bigg[ N_f \dot{\phi}  + 2i \alpha \phi\bigg],\,\,\,\,\,
\pi^\dagger = \frac{\partial L^{(2)}}{\partial \dot{\phi}} = \frac{1}{2}\bigg[ N_f \dot{\phi}^\dagger  - 2i \alpha \phi^\dagger\bigg]
\ea
After the simple canonical transformation
$\pi \rightarrow \sqrt{N_f/2}\, \pi' \, \qquad \phi \rightarrow \sqrt{2/N_f}\,\, \phi'$,
the Hamiltonian
\ba
H^{(2)}[\phi,\pi^\dagger, \pi,\phi^\dagger] &\equiv & \pi^\dagger \dot{\phi} + \pi \dot{\phi}^\dagger - L^{(2)}[\phi,{\phi}^\dagger]
\label{H1}
\ea
becomes
\ba
H^{(2)}[\phi,\pi^\dagger, \pi,\phi^\dagger]  &=&  \pi^\dagger \pi +  ({\cal M}c_0^2)^2 \phi^\dagger \phi - i{\cal M}c_0^2 (\phi \pi^\dagger - \phi^\dagger \pi)+ c_0^2 (\nabla \phi^\dagger)(\nabla \phi) -   (\bar{M}^2/4N_f)(\phi - \phi^\dagger)^2.
\label{H2}
\ea
We diagonalise the Hamiltonian by introducing
new fields $(\Psi, \chi)$ through the canonical transformation
\be
\left( \begin{array}{c}
\chi \\ \Psi^\dagger
\end{array}\right)
=
D \left( \begin{array}{c}
\phi \\ \pi
\end{array}\right)
\qquad \text{where} \qquad
D = \left( \begin{array}{cc}
\sqrt{\frac{{\cal M}c_0^2}{2}} & \frac{i}{\sqrt{2{\cal M}c_0^2}} \\
\sqrt{\frac{{\cal M}c_0^2}{2}} & - \frac{i}{\sqrt{2{\cal M}c_0^2}}  \\
\end{array}\right)
\label{Ann}
\ee
and its conjugate.  That is, $\Psi$ and $\chi$ are the annihilation operators of, respectively, particles and anti-particles in terms of which $\pi^\dagger \pi + ({\cal M}c_0^2)^2 \phi^\dagger \phi= {\cal M}c_0^2(\Psi^\dagger \Psi + \chi^\dagger \chi)$.
On substituting (\ref{Ann}) into the generalised Lagrangian
\ba
L^{(2)}[\phi,\phi^\dagger, \pi,\pi^\dagger] &\equiv & (\dot{\phi}\pi^\dagger + \dot{\phi}^\dagger \pi) -  H^{(2)}[\phi,\pi^\dagger, \phi^\dagger,\pi]
\label{L4}
\ea
where $H^{(2)}$ is given in (\ref{H2}) we find, on approximating $c^2 = {\bar M}_0^2/4\alpha{\cal M}$ that
\ba
L^{(2)} &=&\bigg[\Psi^\dagger \left(  i  \frac{\partial}{\partial t} + \frac{1}{2{\cal M}}  \nabla^2
\right)  {\Psi} + \frac{1}{2}{\cal M}c^2(\Psi - \Psi^\dagger)^2\bigg]
+
\bigg[{\chi}^\dagger \left(  i  \frac{\partial}{\partial t} + \frac{1}{2{\cal M}}  \nabla^2 -2{\cal M}c_0^2
\right)  {{\chi}}  + \frac{1}{2}{\cal M}c^2(\chi - \chi^\dagger)^2\bigg]
\nn
\\
&-&
\frac{1}{2{\cal M}} \left[ \nabla \chi \nabla \Psi^{\dagger} + \nabla \chi^\dagger \nabla {\Psi}\right]
- \frac{1}{2}{\cal M}c^2(\Psi - \Psi^\dagger)(\chi - \chi^\dagger).
\label{line3}
\ea
We shall see that we can interpret the $\Psi,\Psi^\dagger$ fields as annihilating and creating non-relativistic Bogoliubons.
Because of the mixing terms between $\pi$ and $\phi$ in $H_0$ in (\ref{H2}) the rest mass energy of the $\Psi$ field is cancelled by the chemical potential, whereas $\chi$ has rest mass energy (plus chemical potential) $2{\cal M}c_0^2$.

In the BEC regime where $c_0^2 \rightarrow \infty$ this $\chi,\chi^\dagger$ anti-Bogoliubon sector becomes infinitely heavy and decouples.
If we neglect the dynamical degrees of freedom of the anti-Bogoliubons (terms $O(c_0^{-2})$), the action
 \be
S^{(2)} = \int d^4x\,L^{(2)}(\Psi^\dagger, \Psi) \equiv \int d^4x\,\bigg[\Psi^\dagger \left(  i  \frac{\partial}{\partial t} + \frac{1}{2{\cal M}}  \nabla^2 - {\cal M}c^2
\right)  {\Psi} + \frac{1}{2}{\cal M}c^2
\left( \Psi^2 + (\Psi^\dagger)^2 \right)\bigg]
\ee
provides a good approximation close to the BEC limit, when the equations of motion become
\ba
\left(  i  \frac{\partial}{\partial t} + \frac{1}{2{\cal M}}  \nabla^2 - {\cal M}c^2\right)  {\Psi} + {\cal M}c^2\Psi^\dagger &=& 0,
\nonumber
\\
\left(  -i  \frac{\partial}{\partial t} + \frac{1
}{2{\cal M}}  \nabla^2 - {\cal M}c^2\right)  {\Psi^{\dagger}} + {\cal M}c^2\Psi &=& 0,
\label{coupled}
\ea
combined as
\be
\left[ \frac{\partial^2}{\partial t^2} + \left( \frac{1}{2{\cal M}}  \nabla^2 \right)^2
 -  c^2\nabla^2 \right] \Psi = 0,
 \label{xx}
 \ee
That is, the $\Psi$ quanta have the dispersion relation
\be
\omega^2 = c^2 k^2 + \bigg(\frac{k^2}{2{\cal M}}\bigg)^2
\label{bog2}
\ee
where ${\cal M} = (\alpha/\rho_0)M_D$.  In the BEC limit, where $c\to c_{\infty}$, ${\cal M}\rightarrow M_D$,  $\Psi$ is the familiar bogolon.

This identification of $\Psi$ with the bogolon seems odd, since we might have anticipated \cite{bardyn} that the fluctuations $\eta (x) = \varphi(x) - \varphi_0$ of the pairing Feshbach field $\varphi (x)$ of (\ref{Lin}) provided the bogolon in that regime. A further step shows this to be the case. We write the decomposition (\ref{varphi}) as
\be
\varphi(x) = |\varphi_0|(1 + \kappa^{1/2}\epsilon (x)/|\varphi_0|)e^{i\theta(x)}\approx |\varphi_0|(1 + i\sqrt{2}\Phi^{\dagger}(x))
\ee
for small $\epsilon$ and $\theta$
where
\be
\Phi = (\theta + i\kappa^{1/2}\epsilon/|\varphi_0|)/\sqrt{2}.
\label{Phi}
\ee
In the BEC regime for real positive $\varphi_0$, where $\kappa^{1/2}/\varphi_0\rightarrow 1$,
  the pairing field fluctuation $\eta$ can be identified with $\phi$ by
\be
\eta^{\dagger}(x) = -i\sqrt{2}\varphi_0\Phi(x)\rightarrow -i\sqrt{2}\varphi_0\phi(x)
\ee
 That is, up to a constant prefactor the non-relativistic limit field of the pairing field $\eta(x)$ is $\Psi$, the non-relativistic limit field of $\phi$, the bogolon field in this regime.

This identity is only valid in the deep BEC regime. As we move away from it towards the BCS regime it breaks down. Most simply, as $\kappa^{1/2}/|\varphi_0|$ ceases to be unity, the correspondence between the pairing field $\eta$ and $\phi$ falters. 
More importantly, as we move away from the BEC limit the approximations that we made to derive the GP equations (\ref{coupled}) from $\phi$ break down.  There are reasons to be cautious in taking the formalism from deep BEC to deep BCS \cite{lee3} but, whatever the details as we move towards the crossover, $c_0$, which suppresses the anti-particle sector, ceases to be large and it rapidly becomes  impossible to separate out bogolons from their antiparticles. This coupling is enhanced by terms in $\Delta$ that we have omitted, such as the inclusion of $ N_f\Delta(\dot{\phi}^\dagger - \dot{\phi})^2$ in (\ref{newL}).
We note that if $c\rightarrow 0$, then the bogolon is just the dimer in that limit.

 We do not go further than identifying the bogolon quanta, as the interactions are rather complicated in terms of $\theta + i\epsilon$ and not amenable to simple analysis. In order to accommodate interactions, particularly beyond the Bogoliubov approximation, we need a different approach to GP equations.

\section{Gross-Pitaevskii equations for $\theta \pm\epsilon$}

 We have seen the agreement between dimers and simple bosons in the deep BEC regime at the level of bogolons. To go beyond the Bogoliubov approximation requires going beyond the quadratic fluctuations in the density and its derivatives. Let
 \ba
L_{GP}^{0}(\psi,\psi^*) &=& i\psi^{\ast}\dot{\psi} - \frac{1}{2M_D}\nabla\psi^{*}
\cdot\nabla\psi - \frac{1}{2N_f}(|\psi|^2 - \rho^f_0)^2
\nonumber
\\
&=& i\psi^{\ast}\dot{\psi} - \frac{1}{2M_D}\nabla\psi^{*}
\cdot\nabla\psi - \frac{M_Dc_0^2}{2\rho^f_0}(|\psi|^2 - \rho^f_0)^2
\label{GP2B}
\ea
be the GP Lagrangian density for the fermi gas parameters $\rho^f_0$, $N_f$ above or, equivalently, $c^2_0$ (hence the superfix zero in comparison to (\ref{SGPc})). In  the Bogoliubov approximation in the deep BEC regime $N_f\to 0$ it is equivalent to the {\it local} acoustic Lagrangian density (cf. (\ref{Sac}))
 \begin{eqnarray}
 {L}^{0,{\rm ac}}(\theta) &=&  -\frac{1}{2}{\rho^f_0}
   G(\theta) +  \frac{N_f}{4}G^2(\theta).
 \label{LeffU0bis}
 \end{eqnarray}
  Alternatively, $\rho$ becomes an auxiliary field,
 \be
 \rho = \rho^f_0 - N_f G(\theta) + O(N_f^2)
 \ee
 which is another way of saying that the acoustic approximation coincides with the Bogoliubov approximation.
 Defining
 \be
 {\bf v} = {\nabla}\theta/M_D,
 \ee
 we recover the continuity equation
 \be
 \partial_t \rho + \vec{\nabla} \left(\rho \bf{v} \right) = 0.
 \label{cont}
 \ee
In consequence, we can perform an inverse Madelung transformation $\psi = \sqrt{\rho}e^{i\theta}$ from the Lagrangian (\ref{LeffU0bis}) depending only in the single variable $\theta$ to recover the doubly-variabled (\ref{GP2B}).
 That is,
 we can identify
 \begin{equation}
  Z^{0,{\rm{ac}} = \int {\cal D}\theta
  \,e^{i S^0,\rm{ac}}[\theta]}
  \label{ZB0}
 \end{equation}
 with
 \begin{equation}
  Z^0_B = \int {\cal D}\psi{\cal D}\psi^*
  \,e^{i S^0_{GP}[\psi^*,\psi]},
  \label{Z0}
 \end{equation}
 because the latter really has only one degree of freedom in the BEC limit.

Some care is needed since, although $N_f\rightarrow 0$ we retain the second term in (\ref{LeffU0bis}). It simplifies matters to identify those places in the equations in which we can set the BEC limit values without changing the final results as we approach the limit and those where we need to approach the limit carefully. In general, we have to set $N_f\rightarrow 0$ last, after the other limit values have been made.

We follow the tactics used in \cite{aitchison} in describing a BCS model of cold fermi gases. We take $\eta = N_f$, as appropriate for the deep BEC regime as it is for the deep BCS. The BEC Lagrangian density (\ref{Leff}) now becomes
 \be
L_F^{\rm{eff}}(\theta, {\epsilon}) =  \frac{N_f}{4}\ G^2(\theta, {\epsilon}) -\frac{1}{2}{\rho_0^f}
   G(\theta, {\epsilon})
 -{\alpha}{\epsilon}G(\theta, {\epsilon})
 +\frac{N_f}{4}D_t^2({\epsilon},\theta)
  -\frac{1}{4}{\bar M}^2{\epsilon}^2.
  \ee

Again as in \cite{aitchison}, we work with the combinations of modes
\ba
\theta &=& \frac{\theta_+ + \theta_- }{2},\,\,\,\,\,\,
\epsilon = \frac{\theta_+ - \theta_- }{2},
\ea
whence
\be
G(\theta, {\epsilon}) =\frac{1}{2}[G(\theta_+) + G(\theta_-)]\,\,\,\,\rm{and}\,\,\,\,
{\it D}_t({\epsilon},\theta) = \frac{1}{2}[{\it G}(\theta_+) - {\it G}(\theta_-)].
\label{Gthetaepsilon}
\ee
In terms of $\theta_{\pm}$, $L_F^{\rm{eff}}(\theta, {\epsilon})$ becomes
\be
2 L_F^{\rm{eff}}(\theta_+,\theta_-) = \left[L^0_{\rm{ac}}(\theta_+) +  L^{0}_{\rm{ac}}(\theta_-) \right] - \frac{1}{2} {\bar M}^2 \left(  \frac{\theta_+ - \theta_- }{2}  \right)^2
- \alpha (G(\theta_+) + G(\theta_-)) \left(  \frac{\theta_+ - \theta_- }{2}  \right)
\label{2Leff}
\ee
where $L^{0}_{\rm{ac}}(\theta)$ is defined in (\ref{LeffU0bis}) in terms of $N_f$ and $c_0$.

This gives  the partition function
\be
Z_{F} = \int {\cal D}\theta{\cal D}\epsilon\,
\,e^{i{ S}_F^{\rm{eff}}[\theta,\epsilon]}
  \,=
  \int {\cal D}\theta_{+}{\cal D}\theta_{-}\,
\,e^{i{ S}_F^{\rm{eff}}[\theta_+,\theta_-]}
  \label{Zd2}
\ee
From our previous comments we can perform an inverse Madelung transformation in the BEC regime. If we define
\ba
\rho_\pm &=& \rho^f_0 - N_f G(\theta_{\pm}) + \alpha (\theta_+ - \theta_-),
\label{newrhopmdef}
\\
{\bf v}_\pm& =& {\nabla}\theta_\pm/M_D,
\ea
then varying with respect to $\theta_\pm$ gives the equations of motion
\ba
\partial_t \rho_\pm + \vec{\nabla} \left(\rho_\pm  \bf{v}_\pm \right)
&=&  \pm \left\{ \frac{\bar{M}^2 N}{2N_f}(\theta_+ - \theta_- )+ \frac{\alpha}{N_f} \left[2\rho^f_0 -(\rho_+ + \rho_-)\right] \right\}
\label{p3}
\ea
representing continuity equations with sources and sinks.

To reproduce these from a GP action in the limit of $c_0\to\infty$ requires two pseudo-fields
\be
\psi_\pm = \sqrt{\rho_\pm}e^{i \theta_\pm}
\ee
where $\rho_\pm$ and $\theta_\pm$ are considered independent.
A partial inverse Madelung transformation for large $c_0$ gives the Lagrangian density (written in mixed formalism)
\ba
L_{2GP}(\psi_+,\psi^*_+,\psi_-, \psi^*_-)
 &=&[L^{0}_{GP}(\psi_+,\psi^*_+) + L^{0}_{GP}(\psi_-,\psi^*_-) ] - \frac{ \bar{M}^2}{4N_f}N (\theta_+ - \theta_-)^2 - \frac{\alpha}{N_f} \left[2\rho^f_0 -(\rho_+ + \rho_-)\right](\theta_+ - \theta_-)
\label{GPrainD}
\ea
where $L^{0}_{GP}(\psi,\psi^*)$ is defined in (\ref{GP2B}) in terms of $N_f$ and $c_0$.  Varying with respect to $\rho_\pm$ and $\theta_\pm$ gives precisely   the equations
(\ref{newrhopmdef}) and (\ref{p3}) in the limit, in which $\rho_{\pm}$ are auxiliary fields. To convert the interaction Lagrangian density fully to GP fields, we express $\theta_{\pm}$ and $\rho_{\pm}$ in terms of $\psi_{\pm}$ as
\ba
(\theta_+ - \theta_-) &=& \frac{1}{2i} \ln\left(\frac{\psi_+\psi_-^*}{\psi_-\psi_+^*} \right),
\\
\rho_\pm &=& \psi_\pm^* \psi_\pm.
\ea

We have not yet included higher than quadratic powers in $\epsilon$ in anticipation that they are unimportant  in the BEC regime, as we shall see later.  To simplify the discussion we include the effects of the self-interaction $L_D^{\rm eff}$.  Eq.(\ref{2Leff}) contains the term quadratic in the BEC limit $\kappa = |\varphi_0|^2$. The complete Lagrangian density is, again in mixed notation,
\ba
L_{2GP}(\psi_+,\psi^*_+,\psi_-,\psi^*_-)
 &=&[ {L}_{GP}^{0}(\psi_+, \psi_+^*) + {L}_{GP}^{0}(\psi_-, \psi_-^*) ] - \frac{ \bar{M}^2}{4N_f}N (\theta_+ - \theta_-)^2 - \frac{\alpha}{N_f} \left[2\rho^f_0 -(\rho_+ + \rho_-)\right](\theta_+ - \theta_-)
 \nn
 \\
 &-& \frac{1}{12}\bar{M}^2(\theta_+ - \theta_-)^3 - \frac{1}{48}\bar{M}^2(\theta_+ - \theta_-)^4.
\label{GPrainD2}
\ea

Entirely in terms of the fields $\psi_{\pm}$ this becomes
\ba
L_{2GP}(\psi_+,\psi^*_+,\psi_-,\psi^*_-)
 &=&[ {L}^{{0}}_{GP}(\psi_+,\psi^*_+) + {L}^{{0}}_{GP}(\psi_-,\psi^*_-) ] + \frac{ \bar{M}^2}{16 N_f}{N} \bigg[\ln\left(\frac{\psi_+\psi_-^*}{\psi_-\psi_+^*} \right)\bigg]^2
  \\
  \nonumber
  &-& \frac{i\alpha}{2N_f}\left[\psi_+^{\ast}{\psi_+} + \psi_-^{\ast}{\psi_-} - 2\rho^f_0 \right]\ln\left(\frac{\psi_+\psi_-^*}{\psi_-\psi_+^*} \right)
 \nn
 \\
 &-& \frac{i\bar{M}^2}{48\rho^f_0}\,\bigg[\ln\left(\frac{\psi_+\psi_-^*}{\psi_-\psi_+^*} \right)\bigg]^3 - \frac{\bar{M}^2}{192 \rho^f_0} \,\bigg[\ln\left(\frac{\psi_+\psi_-^*}{\psi_-\psi_+^*} \right)\bigg]^4
\label{GPrainD3}
\ea
and we have further used $2|\varphi_0|^2 = \rho^f_0$ in the BEC regime. It is already clear that the last two terms arising from the self-interaction are, relatively, $O(N_f)$, becoming vanishingly small.

In summary, we have shown that the partition function $Z_{2F}$ of (\ref{ZFF}) can, after some circuitous manipulations, be identified with  the partition function
\be
Z_{2GP} = \int {\cal D}\psi_+{\cal D}\psi^*_+{\cal D}\psi_-{\cal D}\psi^*_-
  \,e^{i S_{2GP}[\psi^*_+,\psi_+,\psi^*_-,\psi_-]}
\ee
in the deep BEC regime, where
\be
S_{2GP}[\psi^*_+,\psi_+,\psi^*_-,\psi_-] = \int d^4x\,L_{2GP}(\psi_+,\psi^*_+,\psi_-,\psi^*_-)
\ee
The identification required a derivative expansion, but beyond that it goes further than the Bololiubov approximation, for which the single-field GP action of (\ref{Z}) with action (\ref{SGP}) was sufficient via (\ref{Zd2}), although again only in the deep BEC regime.
Eq.(\ref{GPrainD3}) is the main result of this paper, an effective GP theory in the vicinity of the deep BEC regime.

We note that, if we insert the deep BCS  parameter values $\alpha\to 0, |\varphi_0|\rightarrow 0$ in (\ref{GPrainD3}) we recover the generalised acoustic approximation action of \cite{aitchison} despite the replacement of the Feshbach resonance by an s-wave point atom-atom interaction in their calculations. However,
we cannot extend Eq.(\ref{GPrainD3}) to the intermediate regime  for two very different reasons.  Most simply, the Lagrangian density (\ref{2Leff}) acquires $G(\theta_+)G(\theta_-)$ cross-terms preventing a reconstruction in terms of two GP fields. Further, even in the absence of cross-terms the non-divergence of $c_0$ prohibits inverse Madelung transformations.

 There is a final caveat.  The model that we have considered here describes an effective theory and there is a renormalisation ambiguity due to the higher terms in the Jacobian $J[\epsilon] = {\rm Det}[1  + \kappa^{1/2}\epsilon/|\varphi_0|]$ which cannot necessarily be absorbed in the redefinition of terms \cite{fubini,rivers} and which, by effectively renormalising the self-interaction strength $u_B$, decouples it from $c_{\infty}$. However, they are also ignorable in the deep BEC regime when $N_f \rightarrow  0$.

\section{Static Vortex Solutions}

To demonstrate the difference between the outcomes from the action $S_{2GP}[\psi^*_+,\psi_+,\psi^*_-,\psi_-]$ obtained from (\ref{GPrainD3}) and the GP action $S_{GP}[\psi^*,\psi]$ of (\ref{SGP}) we consider a single vortex as a simple demonstration of collective condensate behaviour.
\\
\\
The existence of vortices in cold fermi gases is confirmed experimentally~\cite{ketterle1,ketterle2,ketterle3} and there has been extensive theoretical analysis of them in the BCS regime~\cite{Nygaard}, the unitary limit of divergent scattering length~\cite{bulgac,machida}, and throughout the BCS-BEC crossover~\cite{chien,ho}, based upon Bogoliubov-de Gennes theory~\cite{strinati,devreese}. In particular, these studies indicate that the  effective Gross-Pitaevskii description
for composite bosons can be provided only in strong coupling of the BEC regime with which, from our very different viewpoint, we agree.

For all the caveats of the previous paragraph, Gross-Pitaevskii theory provides the best way to understand vortices. We have already explored vortex production in the context of the hydrodynamic approximation \cite{lee3} and we return to the problem with the more sophisticated Lagrangian density of Eq.(\ref{GPrainD3}).

For this it is sufficient to fall back onto the GP equations, mean-field equations for the GP partition function, to which we look for non-homogeneous vortex solutions.
We saw that the BEC fermi gas condensate behaved like a condensate of elementary bosons in the hydrodynamic approximation. Such condensates show vortices with global $U(1)$ vortex solutions that are well-understood~{\cite{fetter, kleinert, rorai}}.
Our more realistic system shown above in (\ref{GPrainD3}),
 without this approximation, also permits vortex solutions and we shall contrast them to those of the simpler system. We shall see that, despite the difference in equations, there are no experimentally observable differences in the vortex profile.

 With single vortices in mind, it is sufficient to construct the static limit  of the  Gross-Pitaevskii actions. At the level of approximation that we shall adopt here we insert the deep BEC values $\alpha = \rho_0$ and
$N = N_f + 4 (\rho^f_0)^2/\bar{M}^2$, but do not set $N_f\rightarrow 0$ until calculations have been performed, because of its singular nature.

The static Hamiltonian $H_{GP}$ following from the Lagrangian density $L_{GP}$ of (\ref{GPrainD3}) is then
\ba
H_{GP}(\psi_+,\psi_+^*, \psi_-,\psi_-^*) &=& H^{0}_{GP}({\psi_+,\psi_+^*}) + H^{0}_{GP}({\psi_-,\psi_-^*})
\nonumber
\\
&-&\int d^3x\bigg\{
 \frac{\rho^f_0}{2iN_f} \left[\psi_+^{\ast}{\psi_+} + \psi_-^{\ast}{\psi_-} - 2\rho^f_0 \right]\ln\left(\frac{\psi_+\psi_-^*}{\psi_-\psi_+^*} \right)
\nonumber
\\
&+& \frac{ \bar{M}^2}{16 N_f} N \bigg[\ln\left(\frac{\psi_+\psi_-^*}{\psi_-\psi_+^*} \right)\bigg]^2 \bigg\},
\label{HGP2C}
\ea
where
\ba
H^{0}_{GP}(\psi,\psi^*) &=& \int d^3x\bigg\{ - \psi^{\ast}\bigg[\frac{1}{2M_D}\vec{\nabla}^2 + \frac{ \rho^f_0}{{N}_f}\bigg] \psi +\frac{1}{2{N_f}} (|\psi|^2)^2\bigg\},
\label{HGPH0}
\ea
the Hamiltonian following from the Lagrangian density $L^{0}_{GP}(\psi,\psi^*)$ of (\ref{GP2B}), defined in terms of $N_f$ and $c_0$.
For clarity of equations, initially we have neglected the terms in $(\theta_+ - \theta_-)^3$ and  $(\theta_+ - \theta_-)^4$. We shall include them later.

To see how vortices arise in (\ref{HGP2C}), in cylindrical polars $(r, z, \varphi)$ we make the simple ansatz
\be
\psi_{\pm}(r,\varphi)= f(r) e^{\pm i\epsilon (r)}e^{i\varphi}
\ee
where $f^2 = \rho$. $H_{GP}$ then takes the form (up to an additive constant)
\ba
H_{GP}(f, \epsilon) &=& 2\pi\int r dr\bigg\{  \frac{1}{M_D}\bigg[f'^2  + \epsilon'^2f^2 +  \frac{1}{r^2} f^2\bigg] r
 + \frac{1}{N_f}\left[\rho^f_0 - f^2\right]^2
 \nonumber
\\
&+& \frac{4\rho^f_0}{N_f} \left[\rho^f_0 - f^2\right]\epsilon
+\frac{ \bar{M}^2}{N_f} N \epsilon^2\bigg\}.
\label{HGP3C}
\ea
\begin{figure}
\centering
\includegraphics[width=10.0cm]{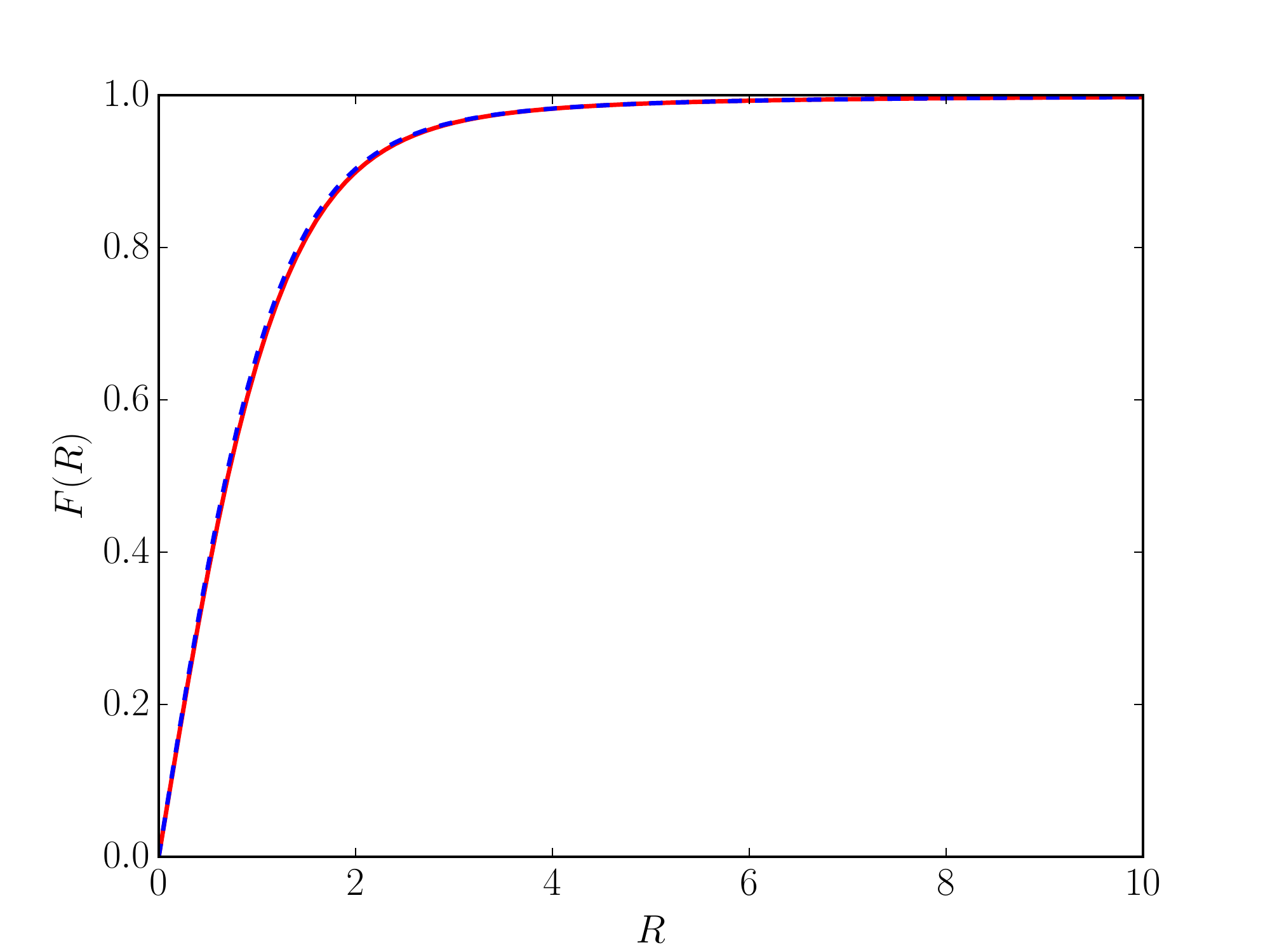}
\caption{
Numerical solution (blue profile) of the BEC vortex Eq.(\ref{VEO3}) compared to the solution to the global U(1) vortex Eq.(\ref{VEO4}) for a condensate of elementary bosons (green profile). The difference is insignificant.
}
\end{figure}
The resulting deep BEC equations of motion are
\ba
  0 &=& f'' + \frac{f'}{r} - \epsilon'^2f -  \frac{1}{r^2}f  +
  2f\frac{M_D}{N_f}[\rho_0 - f^2] + \frac{4M_D\rho^f_0}{N_f}f \epsilon
  \label{VEO}
\\
0 &=& \bigg[\epsilon'' + \frac{1}{r}\epsilon'\bigg]f^2 - \frac{2M_D\rho^f_0}{N_f} \left[\rho^f_0 - f^2\right] - \frac{M_D \bar{M}^2}{N_f} N \epsilon.
\label{VEOM}
\ea
The limit of vanishing $N_f$ in these equations is subtle. We develop iterative solutions in powers of $N_f$ before taking the limit.
The effect is to solve (\ref{VEOM})  as making $\epsilon$ the auxiliary field
\be
\epsilon \approx -\frac{2\rho^f_0}{\bar{M}^2N}(\rho^f_0 - f^2),
\label{epsilon}
\ee
where the approximation reflects the deep BEC limit. Inserting this result into (\ref{VEO})
gives
\be
\epsilon' \approx \frac{4\rho^f_0 f f'}{\bar{M}^2N} = \frac{f f'}{\rho^f_0}
\label{epsilon'}
\ee
in the small $N_f$ limit. Inserting (\ref{epsilon}) and (\ref{epsilon'}) in (\ref{VEO}) gives the modified vortex equation
\be
  0 = f'' + \frac{f'}{r} - f'^2\frac{f^3}{(\rho^f_0)^2} -  \frac{1}{r^2}f  +
  \frac{2M_D^2c^2}{\rho^f_0}f[\rho^f_0 - f^2].
  \label{VEO2}
  \ee
  We see that after some intricate cancellation the singular terms in $1/N_f$ have been replaced by benign factors of $1/N$ or, equivalently, $c^2 = \rho^f_0/M_DN$.  Although, from (\ref{epsilon}), there is a core of density fluctuations, the energy within this cannot be separated from that due to the density profile in (\ref{HGP3C}), again because of the cancelations that are required to give a finite answer. [It is because of the difficulties of implementing these cancelations that other seemingly plausible ansatze for vortices, such as trying to put all the vortex structure into just one of the pseudo-fields (e.g. taking $\psi_- = \rho^{1/2}$) fail.]

  Adding the terms in $\epsilon^3$ and $\epsilon^4$ necessary to get a non-zero $c^2$ in the BEC regime has no effect on the final equations. To see this we observe that the inclusion of the terms due to the self-interaction replaces $H_{GP}$ of (\ref{HGP3C}) by
 \ba
H_{GP}(f, \epsilon) &=& 2\pi\int r dr\bigg\{ -\frac{1}{M_D}\bigg[f f'' - \epsilon'^2f^2 + \frac{1}{r}f f' - \frac{1}{r^2} f^2\bigg] - 2f^2\frac{\rho^f_0}{N_f} + \frac{f^4}{N_f}
\\
&+& \frac{4\rho^f_0}{N_f} \left[\rho_f - f^2\right]\epsilon
+\frac{ \bar{M}^2}{N_f}N \epsilon^2 + \frac{2}{3}\bar{M}^2\epsilon^3 + \frac{1}{6}\bar{M}^2\epsilon^4\bigg\}.
\label{HGP3CC}
\ea
Thus (\ref{VEOM}) becomes
\ba
0 &=& \bigg[\epsilon'' + \frac{1}{r}\epsilon'\bigg]f^2 - \frac{2M_D\rho^f_0}{N_f} \left[\rho^f_0 - f^2\right] - \frac{M_D \bar{M}^2}{N_f}N \epsilon - 2M_D\bar{M}^2\epsilon^2 - \frac{2}{3}M_D\bar{M}^2\epsilon^3.
\label{VEOM2}
\ea
On multiplying through by $N_f$, we see that the additional terms do not contribute as $N_f$ vanishes and (\ref{epsilon}), (\ref{epsilon'}) and (\ref{VEO2}) persist. Higher order terms from the Jacobian would have the same lack of effect.

In particular, the extraction of the bosonic condensate from the fermi pairs does not introduce any new length scales. The equation is cast in dimensionless form by rewriting $f$ as $f\equiv \sqrt{\rho^f_0}F$, in units
  of length $r = \xi R$, where  $\xi = \hbar/M_Dc$ is the length scale of the model. On taking the $N_f\rightarrow 0$ limit into account Eq.(\ref{VEO2}) then becomes
\be
  0 = F''(R) + \frac{F'(R)}{R} - F'(R)^2 F(R)^3 -  \frac{1}{R^2}F(R)  +
  2F(R)[1 - F(R)^2]
  \label{VEO3}
  \ee
 where the primes denote differentiation with respect to $R$ and we have used the fact that $\alpha\approx\rho^f_0$  in the BEC regime.   The boundary conditions are $F(0) = 0, F(\infty) = 1$, whence $F(R)\sim R$, small $R$, as for the simple $U(1)$ global vortex displayed below. The numerical solution is given in Fig.1.

We contrast this result with that obtained from the static Hamiltonian $H^{\rm{ac}}_{GP}$, the Hamiltonian for elementary bosons, derived from the Lagrangian density (\ref{SGP}) after replacing $N_b$ by $N$,
\ba
H^{\rm{ac}}_{GP}(\psi) &=& \int d^3x\bigg\{ - \psi^{\ast}\bigg[\frac{1}{2M}\vec{\nabla}^2 + \frac{ \rho_0}{N}\bigg] \psi +\frac{1}{2N} (|\psi|^2)^2\bigg\}
\label{HGPH}
\ea
where $\psi$ is the single GP field in this approximation.

The ansatz
  \be
  \psi(r,\varphi)= f(r) e^{i\varphi}
  \ee
  leads to the more familiar $U(1)$ vortex equation~{\cite{fetter}},
  \be
  0 = f'' + \frac{f'}{r}  -  \frac{1}{r^2}f  +
  \frac{2M^2c^2}{\rho_0}f[\rho_0 - f^2].
  \label{VEO0}
  \ee
  In identical dimensionless units the equation for
  the conventional global U(1) vortex is~\cite{fetter}
 \be
   0 = F''(R) + \frac{F'(R)}{R}  -  \frac{1}{R^2}F(R)  +
  2F(R)[1 - F(R)^2].
  \label{VEO4}
  \ee
  The difference between (\ref{VEO0}) and the diatomic (\ref{VEO2}) lies  entirely in the $F'(R)^2 F(R)^3$ term, of universal strength in the deep BEC regime.
 The effect of the linear behaviour near the origin is to render the contribution from the $F'(R)^2 F(R)^3$ term negligible, as is seen in the numerical
solutions to Eqs. (\ref{VEO3}) and (\ref{VEO4}) of Fig.1, all but indistinguishable to the naked eye. That is, as far as the vortex profile is concerned it is sufficient to work with the simple GP action of (\ref{GP2B}) for elementary bosons.

Some caution is required in interpreting these results. The original formulation of (\ref{LeffU0}) assumed small fluctuations $\epsilon$. Although
convenient factors of $N_f$ enabled higher powers of $\epsilon$ to occur without having to invoke the smallness of the fluctuations, we are forced to address this in our BEC solution (\ref{epsilon}), which becomes
$\epsilon\approx - (1 - F^2)/2$ whereas, by definition, $\epsilon\rightarrow - (1 - F)$ in the BEC limit. Thus, technically, we should only believe
the vortex solution from(\ref{VEO3}) away from the vortex core where $F\approx 1$. In practice, since $F$ has to vanish at the core, we anticipate the vortex solution of Fig.1 to have greater validity \cite{devreese}.  In this regard there are some similarities with \cite{devreese}, although in our case the density is entirely diatomic.
However, unlike the approach of \cite{devreese}, in our case a new Gross-Pitaevskii action for the BEC regime is constructed with a clear Galilean identification of
 coupled fields as dimers.

\section{Conclusions}

 The main goal of this work was to understand to what extent diatoms/dimers in the BEC regime of a cold fermi gas resembled simple bosons.  We had already observed  \cite{lee3} that, at the level of the acoustic (or hydrodynamic) approximation in which the quasi-particle phonon has dispersion relation $\omega = ck$, the condensate of a cold fermi gas formally mimics that of simple bosons across the whole BEC-BCS regime. By that we mean that the path integral partition functions of the two condensates are identical (in each case the Gross-Pitaevskii partition function). Since we know that the GP partition function is that for the Grand Canonical ensemble of simple bosons the identity is established at that level. We are careful in preserving Galilean invariance at each step.

 We have pursued this is two stages. As our first step we restrict ourselves to the Bogoliubov approximation, where we have shown that in the deep BEC regime the dimer and elementary boson condensates  are identical (from the point of view of partition functions).
 Further, in this regime we identify the bogolon, with dispersion relation
 \be
 \omega^2 = c^2k^2 + (k^2/2M_D)^2
 \ee
 as the non(sound-cone)-relativistic quantum excitation of the fluctuation field. For this identity an essential ingredient in the comparison is the divergence of the mean-field speed of sound $c_0$ (whereas the true speed of sound $c$ remains finite). Outside the deep BEC regime the relationship fails and a description in terms of a single Gross-Pitaevskii field is not possible.

 Our second step is to go beyond the Bogoliubov approximation in the BEC regime. To do this we have developed a generalised Gross-Pitaevskii action for {\it two} coupled fields which describes the dimer dynamics. We stress that these identities take place at the level of partition functions and are not mean-field results.
The two-field action differs from the conventional GP action. As a simple demonstration as to what this difference may mean we contrast the form of a single linear vortex for the dual-field fermi gas to that of the conventional GP action.
 Although the vortex equations are universally different  in the deep BEC regime they turn out to be almost indistinguishable numerically.  This near-identity of the solutions helps confirm the supposition that the strongly-coupled dimers of the  BEC regime do behave like elementary bosons to a good approximation, even beyond the Bogoliubov picture.

We stress that the effective {\it time-dependent}  Gross-Pitaevskii theory in the BEC regime can, in principle, be implemented beyond the static approximation to study the dynamical aspects of vortices, something that is difficult from the Bogoliubov- de Gennes perspective.

As we move away from the deep BEC region towards the BCS regime the situation becomes more complicated for two very different reasons.  Most simply, with $N_f\neq\eta$ no longer small the Lagrangian density (\ref{2Leff}) no longer separates dynamically and finiteness of $c_0$ prevents inverse Madelung transformations to GP fields, even if it were possible.

\section{Acknowledgements}
R.J.R would like to thank APC, U. of Paris 7, for hospitality, where some of this work was performed.
This work of D.S.L. was supported in part by the
Ministry of Science and Technology, Taiwan, and the short-term visiting program in Academia Sinica (Taiwan).

\section{Appendix}
In this Appendix, we provide detailed expressions of the coefficients in the effective action~(\ref{Leff}) allowing for a self-interaction term $L_D^{\bf eff} = -u_B|\varphi|^4/4$.

The  chemical potential $\mu$ and  $|\varphi_0|$ are defined through
\be
\int \frac{d^3{\bf p}}{(2\pi)^3}
\bigg( \frac{1}{2E_p} - \frac{1}{2\epsilon_{p}} \bigg)  - \frac{u_B}{2g^2}|\varphi_0|^2 = \frac{\nu - 2\mu}{g^2}
\equiv -\frac{m_f}{4\pi a_S}
\label{gap2}
\ee
and the fermion number density $\rho_0^f$,
\begin{equation}
 \rho_0^f = \rho^{\rm{ex}}_0 +
2\rho^D_0 ,
\end{equation}
  where
\begin{equation}
 \rho^{\rm{ex}}_0 = \int \frac{d^3 {\bf p} } { (2\pi)^3 } \ \left[ 1 -
\frac{\varepsilon_{p}}{E_{p}}
  \right]
\end{equation}
is the explicit fermion density and $\rho^D_0 = |\varphi_{0}|^2$.
\\
\\
  In conventional notation, $\varepsilon_{p}=p^2/2m_f-\mu$ and $E_{p}=(
\varepsilon_{p}^2 + g^2|\varphi_{0}|^2 )^{1/2} $.
The principal coefficients take the form (after renormalisation  \cite{lee1})
\begin{eqnarray}
N_f &=& g^2 |\varphi_0|^2\int \frac{d^{3} {\bf p}}{(2\pi)^3} \frac{1}{ 2 E_{p}^3},\,\,\,\,\, \,\,\,\,\,
\alpha= 2|\varphi_{0}|
\kappa^{1/2}
\bigg[1 + \frac{1}{2}g^2\int \frac{d^3 {\bf p}}{(2\pi)^3}
\frac{\varepsilon_{p}}{ 2E_{p}^3}\bigg] \, ,
\\
 {\bar M}^2 &=& 4 \kappa (\nu-2 \mu)-2 g^2 \kappa \int
\frac{d^3 {\bf p}}{ (2\pi)^3} \bigg[\frac{\varepsilon_{p}^2}{
E_{p}^3} - \frac{1}{\epsilon_p}\bigg],\,\,\,\,\,\,\,   \,
\eta = g^2\kappa \int \frac{d^3 {\bf p}}{(2\pi)^3}
\frac{\varepsilon^2_{p}}{ 2 E_{p}^5} \,
\label{defs}
\end{eqnarray}
with $\epsilon_p=p^2/2m_f$.
The scale factor $\kappa$ is defined as
\be
\kappa \;=\;
\frac{\rho_0}{4m_f g^2\zeta + 2} \, ,
\ee
where
\be
\zeta =  \int \frac{d^3 {\bf p}}{
(2\pi)^3}\bigg[\frac{1}{8 E_{p}^3} \bigg[ \bigg( 1-3 \frac{g^2
|\varphi_0|^2}{E_{p}^2}\bigg) \frac{\varepsilon_{p}}{m_f}
+ \bigg( 5 \frac{g^2 |\varphi_0|^2}{E_{p}^2} \bigg( 1-\frac{g^2
|\phi_0|^2}{E_{p}^2}\bigg) \bigg) \frac{|{\bf p}|^2 \, ({\hat {\bf
p}} \cdot {\hat \nabla})^2 } {m_f^2} \bigg] \bigg] \, .
\label{zeta}
\ee
In (\ref{zeta}) $ {\hat {\bf p}}$ and  ${\hat \nabla}$ are the unit
vectors along the direction ${\bf p}$ and the direction of the
spatial variation of the phase mode $\theta$ respectively.

The inclusion of a self-interaction term $L_D^{\bf eff} = -u_B|\varphi|^4/4$ replaces ${\bar M}_0^2$ by
\be
{\bar M}^2 = {\bar M}_0^2 +6\kappa u_B|\varphi_0|^4.
\ee
as discussed in the text but otherwise there is no explicit $u_B$ dependence.

Formally there is great simplification in the deep BCS and BEC regimes as the expressions above stand. The chemical potential in the deep BCS regime is $\mu=\epsilon_F$, turning
negative as $\vert \mu \vert \simeq 1/2 m_f a_S^2 \rightarrow \infty$ in the deep BEC regime.
The mode-coupling coefficient $\alpha$  vanishes in the deep BCS regime as a consequence of particle-hole symmetry whereas, in the deep BEC regime the scale factor $\kappa$ behaves as
$\kappa \approx \rho_0/2$ so that $\alpha \rightarrow \rho_0$.

In the BEC regime both $N_f$ and $\eta_0 \approx N_f \simeq g^2 \vert \varphi_0 \vert^2 (2 m_f \vert \mu \vert)^{3/2}/ \vert \mu \vert^3 $ vanish due to $\vert \mu \vert \rightarrow \infty$, with $N_f/\eta\rightarrow 1$, again independently of $u_B$.

Further,
in the definition of $\bar
M_0^2$, we have used the relationship between the $s$-wave scattering
length $a_S$ and the binding energy in~(\ref{gap}).
In the deep BEC regime $\bar
M_0$ vanishes as $N_f$ such that $\bar
M_0^2/N_f \rightarrow 2 g^2 \rho_0$.

For completeness, in the deep BCS regime $ N_f $ is the density of states at the Fermi surface, namely $N_f = m k_F/2 \pi^2$. Then in this regime the scale factor $\kappa$ is approximated by
$\kappa \simeq \rho_0/4 m_f g^2 \zeta$  in that  $\zeta \approx k_F \epsilon_F/ 18 \pi^2 g^2 \vert \varphi_0 \vert^2$.
With the approximated results of $\kappa$ and $\zeta$, the behaviour of  $\eta$ in the deep BCS regime is simply given by $\eta \simeq 3 \rho_0/4 \epsilon_F =m_f k_F/ 2 \pi^2 $ when $\rho_0 \equiv k_F^3/3 \pi^2 $ is defined.  Thus, $N_f \approx \eta$ in the deep BCS regime.

\subsection{Numerical values}

It is convenient to work with the dimensionless couplings $\bar g$ and ${\bar u}_B$,
 defined by ${\bar g}^2 = (3 k_F^3/64\epsilon^2_F)g^2$ and ${\bar u}_B =(3/64) (\epsilon_F/k_F)u_B $. To be concrete, consider the resonance in $^6Li$ at $H_0 = 543.25 G$, discussed in some detail in \cite{strecker}. [This is
to be distinguished from the very broad Feshbach resonance
in $^6Li$ at 850 G.]  As our benchmark we take the number density $\rho_0\approx  3 \times 10^{12} cm^{-3}$, whence $\epsilon_F\approx  7 \times 10^{-11} eV$, ${\bar g} = 0.9$. For exemplary purposes we take ${\bar u}_B = 0, 1.5$ and $3.0$. We see that, essentially apart from the crucial $c^2$, there is no dependence on $u_B$ in the deep BEC regime.

\begin{figure}
\centering
\includegraphics[width=16.0cm]{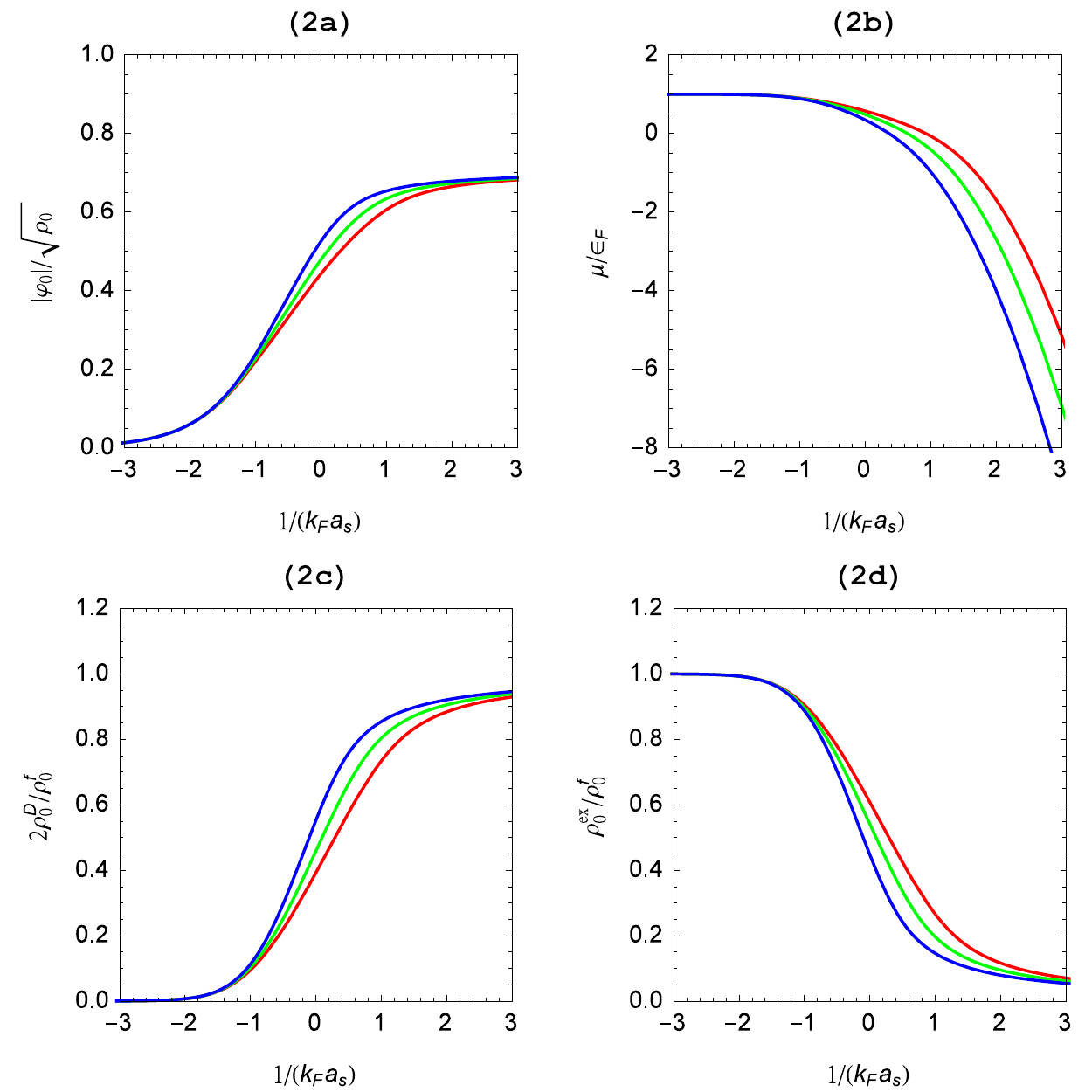}
\caption{
The plots of $\mu$, $\mid\varphi_0\mid$, $\rho_B$, and $\rho_F$ as functions of $1/k_Fa_S$ for the value ${\bar g}=0.9$ and
${\bar u}_B=0(blue),1.5(green),3(red)$.
}
\end{figure}
\begin{figure}
\centering
\includegraphics[width=16.0cm]{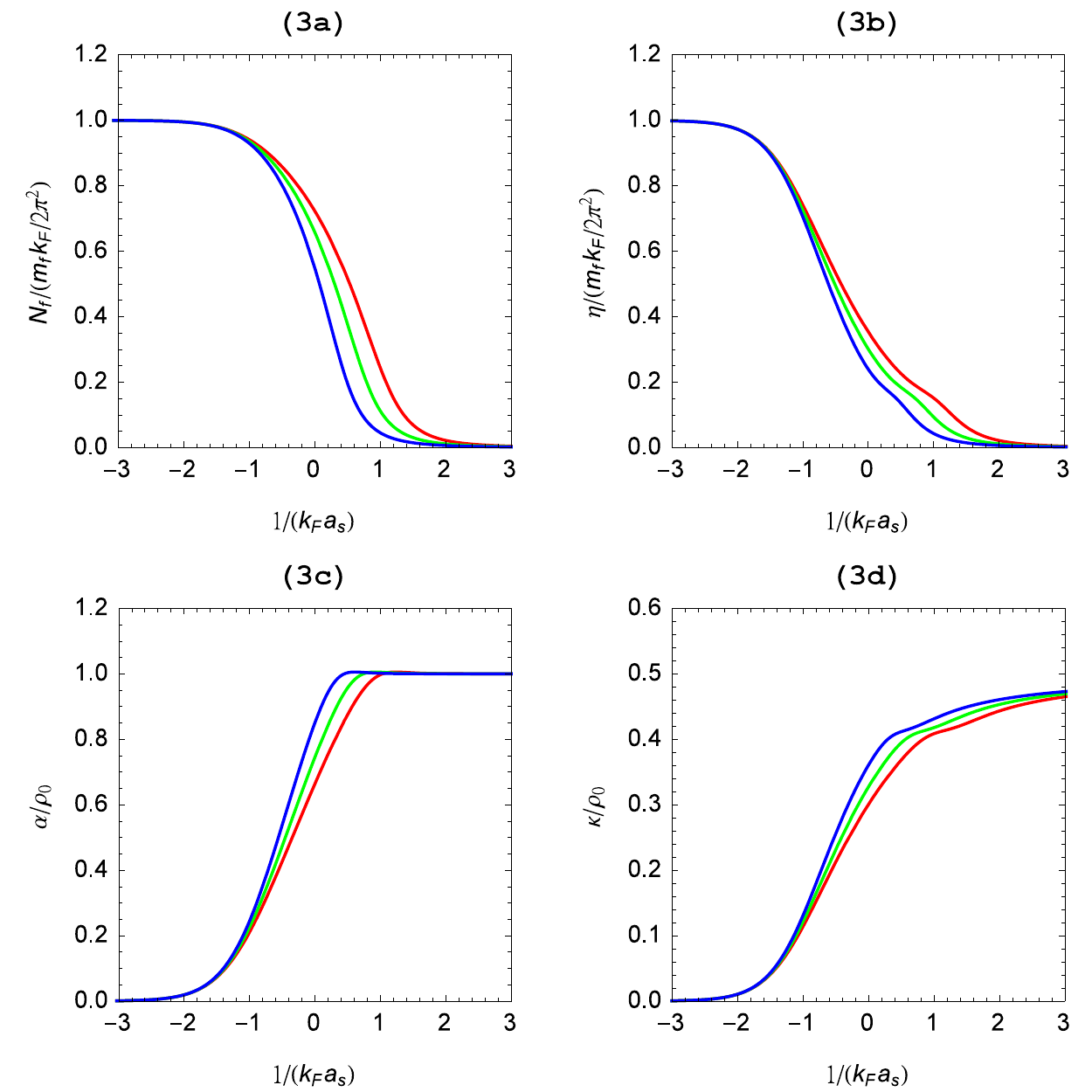}
\caption{
The plots of $ N_0/(mk_F/2\pi^2)$, $\eta/(mk_F/2\pi^2)$, $\alpha/\rho_0$, and $\kappa/\rho_0$ as functions of $1/k_Fa_S$ for the value ${\bar g}=0.9$ and
${\bar u}_B=0(blue),1.5(green),3(red)$.
}
\end{figure}
\begin{figure}
\centering
\includegraphics[width=16.0cm]{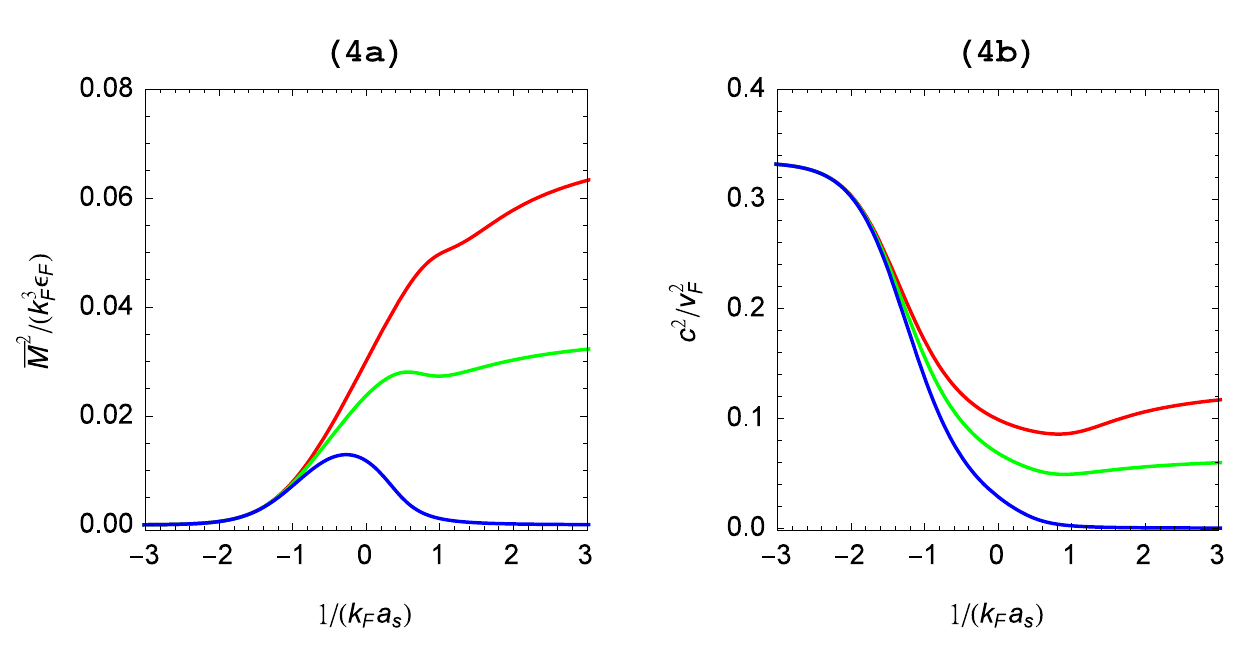}
\caption{The plots of $ {\bar M}^2/(p_F^3 \epsilon_F)$ and $c^2/v_F^2$ as a function of $1/k_Fa_S$ for the value ${\bar g}=0.9$ and
${\bar u}_B=0(blue),1.5(green),3(red)$. In the BCS regime $c^2/v_F^2\rightarrow 1/3$.
 }
\end{figure}


\end{document}